\documentstyle{mn}

\title{A new numerical approach to the oscillation modes of relativistic stars}

\author[Nils Andersson, Kostas D. Kokkotas and Bernard F. Schutz]
{ Nils Andersson$^{1}$, Kostas D. Kokkotas$^{2}$ and Bernard F. Schutz$^{1}$ \\
$^{1}$ Department of Physics and Astronomy,
University of Wales College of Cardiff, PO Box 913,
Cardiff CF2 3YB, United Kingdom \\
$^{2}$ Department of Physics, Aristotle University of Thessaloniki,
Thessaloniki 54006, Greece}

\date{Accepted 1995  (?).
      Received 1994  (?);
      in original form  1994}

\begin{document}

\maketitle

\begin{abstract}
The oscillation modes of a simple polytropic stellar model are
studied.  Using a new numerical approach (based on integration for
complex coordinates) to the problem for the stellar exterior we have
computed the eigenfrequencies of the highly damped w-modes. The results
obtained agree well with recent ones of Leins, Nollert and Soffel [{\em
Phys. Rev. D} {\bf 48} 3467 (1993)].  Specifically, we are able to
explain why several modes in this regime of the complex frequency plane
could not be identified within the WKB approach of Kokkotas and Schutz
[{\em Mon. Not. R. astr. Soc.} {\bf 255} 119 (1992)]. Furthermore, we
have established that the ``kink'' that was a prominent feature of the
spectra of Kokkotas and Schutz, but did not appear in the results of
Leins {\em et al.}, was a numerical artefact. Using our new numerical
code we are also able to compute, for the first time, several of
the slowly damped (p) modes
for the considered stellar models. For very compact stars we find,
somewhat surprisingly, that the damping of these modes does not
decrease monotonically as one proceeds to higher oscillation
frequencies. The existence of low-order modes that damp away much
faster than anticipated may have implications for questions regarding
stellar stability and the lifetime of gravitational-wave sources.
The present results illustrate the accuracy and reliability of the
complex-coordinate method and indicate that the method could prove to be
of great use also in problems involving rotating
stars. There is no apparent reason why the complex-coordinate approach
should not extend to rotating stars, whereas it is accepted that all
previous methods will fail to do so.

\end{abstract}

\begin{keywords}
Stars : neutron stars - Gravitational waves
\end{keywords}

\section{Introduction}

In the concluding remarks of their pioneering paper on nonradial
oscillations of neutron stars Thorne and Campollataro
\shortcite{thorne67} described it as ``{\em just a modest introduction
to a story which promises to be long, complicated and fascinating}''.
The story has undoubtedly proved to be intriguing, and many
authors have contributed to our present understanding of the
pulsations of neutron stars.  Of special interest are the
attempts to calculate the eigenfrequencies of the so-called quasinormal
modes of a stellar system. These are solutions to the perturbation
equations which agree with the necessary boundary conditions at the
centre and at the surface of the star, and at the same time behave as
purely outgoing waves at spatial infinity. In general relativity
pulsations of the stellar fluid will normally be damped due to the emission of
gravitational radiation. Hence, the characteristic frequencies of the
system will have complex values. The real parts correspond to  physical
oscillation frequencies, while the imaginary parts describe the damping
rate of each oscillation mode.

The first attempts to compute outgoing-wave modes for relativistic
stars concerned modes that have an analogue in  Newtonian theory. It
was believed (and verified by the early calculations for nonrotating
stars) that the Newtonian modes of oscillation are shifted only
slightly because of the coupling to gravitational radiation: Each
characteristic frequency adopts a very small imaginary part. In
practice, it is not straightforward to determine these modes with great
accuracy. This difficulty became obvious with the first numerical
calculations \cite{thorne69}. An approach based on a variational
principle devised by Detweiler and Ipser \shortcite{detweiler73} also
suffers from this problem \cite{detweiler75}.  The difficulty of
determining small imaginary parts with some accuracy was not overcome
until much later when Lindblom and Detweiler \shortcite{lindblom83}
combined numerical integration of the equations for the interior with a
numerically integrated solution for the exterior.  Using this technique
they computed the fundamental (f) mode frequency for many realistic
neutron star equations of state. However, they later realized that the
fourth-order system of equations they had used for the stellar interior could
become singular. Consequently, they suggested  that a different choice
of dependent variables be made \cite{detweiler85}. Also worth
mentioning in this context is a paper by Balbinski {\em et al.}
\shortcite{balbinski85} where results obtained by the
Lindblom-Detweiler scheme were found to be in satisfactory agreement
with the  predictions of the standard
quadrupole formula.

The last few years have seen several important contributions to this
field. In an impressive series of papers, Chandrasekhar and Ferrari
have reformulated the problem. In their final system of equations all
fluid perturbations have been eliminated \cite{chandra91a}. This has
the advantage that one can view the problem of gravitational waves
being scattered by a star in a way similar to that used for black holes
\cite{chandra83}. In the second  paper of the series, Chandrasekhar and
Ferrari show that odd-parity (axial) modes can be excited by the
even-parity (polar) modes in the slow rotation limit \cite{chandra91b}.
The coupling is due to the Lense-Thirring effect. In the following
papers, they demonstrate that their algorithm for computing
eigenfrequencies can be trusted so long as the imaginary part is
considerably smaller than the real part \cite{chandra91c}, that the
odd-parity (axial) modes can be of importance for extremely compact
stars \cite{chandra91d} and that the idea of a complex angular momentum
can be useful in astrophysical situations \cite{chandra92}. A summary
of their work has been written by Ferrari \shortcite{ferrari92}.
Another recent reformulation of the equations governing a perturbed
stellar model was proposed by Ipser and Price \shortcite{ipser91}.
They discuss the relation between the Regge-Wheeler gauge (and the
resulting equations) and the diagonal gauge used by Chandrasekhar and Ferrari
\cite{price91}. In fact, they reformulate  the stellar perturbation
equations in such a way that all fluid perturbations can be eliminated
also  in the Regge-Wheeler gauge.  An interesting question is whether
these new formulations of the pulsation problem will make computation
of mode-frequencies any easier. We will not approach that issue here,
but feel that an investigation of that kind is of great importance and
should be encouraged.

Another chapter of the story begins with the study of a  very simple
toy model. Kokkotas and Schutz \shortcite{kokkotas86} suggested that a
plausible model for a stellar system would be a finite string  --
representing the fluid of the star -- coupled by means of a spring to a
semi-infinite string -- the dynamical spacetime. They found that such
systems could accommodate a new family of oscillation modes. These
would be associated with the gravitational degrees of freedom, rather
than the pulsations of the fluid, and would be strongly damped.
Recently, a slightly more realistic toy model led Baumgarte and Schmidt
\shortcite{baumgarte93} to much the same conclusions.  Yet another toy
model, suggested by Kokkotas \shortcite{kokkotas85}, shows a lot of
similarities with the axial (odd parity) modes discussed by
Chandrasekhar and Ferrari \shortcite{chandra91c} and has recently led
to the discovery of a branch of strongly damped axial modes
\cite{kokkotas94}.

A computation of highly damped modes for a realistic stellar model is
not trivial, however. The main difficulty involves numerically
separating the ingoing and outgoing-wave solutions at spatial infinity:
The ingoing solution dies exponentially as $r\to \infty$ while the
outgoing one grows. This problem is well-known from studies of
quasinormal modes for black holes (see Andersson {\em et al.}
\shortcite{andersson93} for a detailed discussion). It seems reasonable
to assume that methods that have proved useful in studies of the
black-hole problem can be adapted to the stellar situation.  A series
expansion approach based on a four-term recurrence relation -- similar
to that developed by Leaver for black holes \cite{leaver85} -- was used
by Kojima \shortcite{kojima88} to verify that the strongly damped
modes  (referred to as w-modes because of their connection to the
gravitational \underline{w}aves) indeed do exist for realistic stellar
models. Kokkotas and Schutz \shortcite{kokkotas92} used a WKB approach
(essentially a geometrical optics assumption of no reflection of waves
in the exterior vacuum) in their calculations of  w-mode spectra for
several
 models.  Their main results have recently been verified by Leins, Nollert
 and Soffel \shortcite{leins93}. In their calculations, Leins {\em et
 al.} employed two different approaches for the exterior: Leaver's
continued fraction approach \cite{leaver85} and a Wronskian technique
that has proved extremely powerful for Schwarzschild black holes
\cite{nollert92}.

Although the results obtained by Leins {\em et al.}  generally  agree
with those of Kokkotas and Schutz there are some differences. Leins
{\em et al.} found  new modes with considerably smaller oscillation
frequencies and also higher damping than those that had been found by
the WKB-technique.  Their calculations also suggest that a ``kink''
that was apparent in the spectra of Kokkotas and Schutz does not exist.
At first sight, one would be tempted to believe that the calculations
of Leins {\em et al.} are the most reliable since the two approaches
they used to deal with the exterior do not, in principle, involve any
approximations. This, however, does not explain why the WKB approach
becomes less reliable as the oscillation frequency increases. In fact,
this is contrary to all expectations: The geometrical optics argument
should be valid for high frequencies.

In this paper we attempt to settle these matters. We will combine a
numerical phase-amplitude approach (similar to that used by Andersson
\shortcite{andersson92} for black holes) to the exterior problem with
the Lindblom-Detweiler  \cite{detweiler85} scheme for the inside. The key idea
is to separate ingoing and outgoing solutions by numerically calculating their
analytic continuations to a place in the complex-coordinate plane where
they have comparabe amplitudes.
This is a new approach that could prove to be of great importance,
especially for problems involving rotating stars.
In such problems the exterior spacetime is not known analytically
and previous methods (such as the WKB method of Kokkotas and Schutz
\shortcite{kokkotas92} and the two methods employed
by Leins {\em et al.} \shortcite {leins93}) will consequently fail.
But it seems likely that a method based on complexifying the coordinates
will work even if the spacetime itself can only be approximated.
In fact, the method proposed  here may well prove to be the only one that
remains useful for the strongly damped
modes of rotating stars. The present
approach also has the advantage that, although based on numerical
integration -- and in that sense it provides an arbitrarily high
numerical precision -- it solves for quantities that are directly
comparable to those used in the WKB scheme of Kokkotas and Schutz.
Hence, in this first application of the new method, we hope to understand
the reason for the discrepancies between
the results of Kokkotas and Schutz \shortcite{kokkotas92} and Leins
{\em et al.} \shortcite{leins93} and thus contribute  further to the
understanding of the oscillation modes of nonrotating relativistic stars.

\section{Nonradial pulsations of a relativistic star}

In Regge-Wheeler gauge the perturbed metric for a relativistic
stellar model takes the form \cite{detweiler85}
\begin{eqnarray}
ds^2 &=& -e^\nu \left(1+r^\ell H_0 e^{i\omega t}Y_{\ell m}\right)dt^2 -
\nonumber \\ &&
 2i\omega r^{\ell+1} H_1 e^{i\omega t}Y_{\ell m} dt dr +
\nonumber \\&&
e^\lambda \left(1-r^\ell H_0 e^{i\omega t}Y_{\ell m}\right)dr^2 +
\nonumber \\&&
r^2\left( 1-r^\ell K e^{i\omega t}Y_{\ell m}\right)\left(d\theta^2 +
\sin^2 \theta d\phi^2\right) \ ,
\label{metric}\end{eqnarray}
where $H_0$, $H_1$ and $K$ are functions of $r$ only, since we have assumed
 a harmonic dependence on time. We also have the standard definition of
$ M(r) $:
\begin{equation}
e^{-\lambda} = 1 - {2 M \over r} \ ,
\end{equation}
which plays the role of an effective mass inside radius $r$.

As is well-known, a stellar model in equilibrium is described by the
Tolman-Oppenheimer-Volkov equations [see for example equations
(A3)-(A5) of Lindblom and Detweiler \shortcite{lindblom83}]. These are
a system of three coupled first order differential equations  that
determine the  mass  $M$, the metric quantity $\nu$ and the
pressure $p$. Their solution requires that an equation of state $\rho =
\rho(p)$ for the energy density is specified. In the present
investigation we will assume a simple polytropic equation of state
\begin{equation}
p = \kappa \rho^2 \ ,
\end{equation}
where $\kappa = 100 {\rm km}^2$ and we are using units in which $c=G=1$
throughout this paper.
This choice of equation of state may not be very realistic, but it
simplifies the calculations
somewhat. It also enables us to compare our results
directly to those of Balbinski {\em et al.} \shortcite{balbinski85}, Kojima
\shortcite{kojima88}, Kokkotas and Schutz \shortcite{kokkotas92} and
Leins {\em et al.} \shortcite{leins93}.

Inside the star, the perturbed fluid is described by a Lagrangian
displacement $\xi^a$, where
\begin{equation}
\xi^r = r^{\ell-1}e^{-\lambda/2}W e^{i\omega t}Y_{\ell m} \ ,
\end{equation}
\begin{equation}
\xi^\theta = -r^{\ell-2} V e^{i\omega t}{\partial \over \partial
\theta}Y_{\ell m}\ ,
\end{equation}
\begin{equation}
\xi^\phi = - {r^{\ell-2} \over \sin^2 \theta} V e^{i\omega t}{\partial
\over \partial \phi}Y_{\ell m}\ .
\end{equation}

On general grounds one  would think that perturbations of a spherical
star have four
degrees of freedom. Two of these are associated with the fluid and the
remaining two correspond to the gravitational waves. This means that
the five functions $H_0$, $H_1$, $K$, $W$ and $V$ should not be independent.
As was shown by Detweiler and Lindblom
\shortcite{detweiler85} a system of equations that is free from
singularities can be obtained by defining
\begin{eqnarray}
&&X = \omega^2 (\rho+p) e^{-\nu/2}V-r^{-1} p^\prime e^{(\nu-\lambda )/2}W
+ \nonumber \\
&& {1\over 2}(\rho+p)e^{\nu/2}H_0 \ ,
\end{eqnarray}
where a prime denotes a derivative with respect to $r$.
(A misprint in equation (6) of Detweiler and Lindblom
\shortcite{detweiler85} has been corrected here.)
Then it follows, as a consequence of Einstein's equations, that
\begin{eqnarray}
&&(2M +nr+Q)H_0 = 8\pi r^3e^{-\nu/2}X - \nonumber \\
&& \left[ (n+1)Q -\omega^2r^3
e^{-(\nu+\lambda)} \right] H_1  + \nonumber \\
&& \left[ nr -\omega^2r^3e^{-\nu} - {e^\lambda \over r} Q(2M - r +Q)
\right] K \ ,
\end{eqnarray}
where
\begin{equation}
Q = M + 4\pi r^3 p \ ,
\end{equation}
and
\begin{equation}
n = {1\over 2} (\ell + 2) (\ell -1) \ .
\end{equation}

The interior problem now reduces to a system of four first order
differential equations for $H_1$, $K$, $W$ and $X$ [equations (8)-(11)
in Detweiler \& Lindblom \shortcite{detweiler85}].
Only two of the four linearly independent solutions to this system
are well-behaved at the centre of the star (at $r=0$). Furthermore, the
perturbed pressure must vanish at the surface ($r=R)$, which
implies that $X(R) = 0$.
These conditions specify a single acceptable solution for each frequency
$\omega$. Physically, this solution describes the response of the star
when gravitational waves of the given frequency are incident upon it.

In the exterior of the star the fluid perturbations vanish, and the two
metric perturbations $H_1$ and $K$ can be combined in such a way
[see for example Fackerell \shortcite{fackerell71}]
that one obtains a single second-order differential
equation known as the Zerilli equation. This equation can be written
\begin{equation}
\left[ {d^2 \over dr_\ast^2} + \omega^2 - V_Z(r) \right] Z = 0 \ ,
\label{zerilli}\end{equation}
with the effective potential $V_Z$ given by
\begin{eqnarray}
&&V_Z(r) = 2 \left( 1 - {2M\over r} \right) \times \nonumber \\
&& {n^2(n+1)r^3 +
3n^2Mr^2 +
9nM^2r + 9M^3 \over r^3(nr+3M)^2 } \ .
\end{eqnarray}
The tortoise coordinate $r_\ast$ is defined by
\begin{equation}
{d \over dr_\ast } = \left( 1 -{2M \over r} \right)  {d \over dr} \ .
\label{tortoise}\end{equation}
Clearly, this equation allows two linearly independent solutions. Far
away from the star one of these can be identified as an outgoing wave,
whereas the other describes an ingoing wave. For a general frequency,
the physically acceptable  solution for the interior of the star
leads to a mixture
of out- and ingoing waves at spatial infinity. The
quasinormal modes of the system are those special frequencies
for which no waves come in from infinity. In a sense, these are the
 frequencies at which the star can be expected to radiate ``spontaneously''.

\section{The phase-amplitude approach}

In any numerical approach it is imperative that the quantities under
consideration are smooth and nicely behaving. If that is not the case,
an integration scheme may require ridiculously small steps in order to
achieve the desired accuracy. One standard way to avoid this difficulty
when one has to solve for (say) an oscillating function -- such as
the wave-like solutions in the stellar problem  -- is to appeal to
general properties of the solutions and express them in terms of slowly
varying functions. In the following we adopt this approach to deal
with the exterior problem for stars.

Although the tortoise coordinate $r_\ast$ appears naturally
[cf. (\ref{zerilli})] in a
formal description of the  perturbation equations it may obscure a
numerical solution of the problem, especially as we plan to use
complex coordinates. In order to avoid any difficulties we introduce a new
dependent variable $\Psi$ according to
\begin{equation}
Z = \left( 1 - {2 M \over r} \right)^{-1/2} \Psi \ .
\label{zpsi}\end{equation}
Then we get a new differential equation \cite{andersson93}
\begin{equation}
\left[ {d^2 \over dr^2} + U(r) \right] \Psi = 0 ,
\label{ode}\end{equation}
where
\begin{equation}
U(r) = \left( 1 - {2 M \over r} \right)^{-2} \left[
\omega^2 - V_Z(r) + {2M\over r^3} -  {3M^2
\over r^4} \right] \ .
\end{equation}

One of the important properties of an equation such as (\ref{ode}) is
that  the Wronskian of any two linearly independent solutions must be a
constant. This means that the general solution can always be
constructed as a combination of two basic solutions that has the form
\begin{equation}
\psi^\pm = q^{-1/2} \exp \left[ \pm i \int q dr \right] \ ,
\label{pamsols}\end{equation}
where the function $q(r)$ is a solution to the nonlinear equation
\begin{equation}
{1\over 2q} {d^2q \over dr^2} - {3 \over 4q^2} \left( {dq \over dr}
\right)^2
+q^2 - U = 0 \ .
\label{qequation}\end{equation}

At first sight, it may seem as if we have replaced the relatively
simple second-order differential equation (\ref{ode}) with a more
complicated equation. In principle that may be true, but it should be
quite obvious that the function $q$ -- as determined from
(\ref{qequation}) -- may be slowly varying even when the solution to
(\ref{ode}) oscillates wildly.

Moreover, it is easy to make a connection to the WKB approximation
here. Whenever $U$ is a slowly varying function of $r$ one can neglect
the two derivatives in (\ref{qequation}). That is, one can use $q \approx
U^{1/2}$. Since that is the case for large $r$, we shall use this
approximation to generate initial values for the integration of
(\ref{qequation}) far away from the star. This is done in such a way
that $q \to \omega$ as $r\to +\infty$, {\em i.e.}, the function $\psi^+$
represents an ingoing wave whereas $\psi^-$ is outgoing.

\section{Stokes phenomenon - Why the WKB approach breaks down}

The purpose of this section is to study to what extent a WKB
approximation of the Zerilli function at the surface of the star can be
trusted. Specifically, we want to investigate whether a geometrical
optics argument that the waves undergo no reflection in the exterior
spacetime [as was assumed by Kokkotas and Schutz \shortcite{kokkotas92}] can
be used in this problem. If that is the case, one can simply express
the exterior solution at the surface in terms of $U^{1/2}(R)$.
In general, one would expect this kind of argument to be valid
for high frequencies. In the case of really low frequencies, however,
the situation becomes more obscure. Then waves can possibly be
backscattered by the curvature of spacetime.

These qualitative expectations
agree with  the results of Leins {\em et al.} \shortcite{leins93}
who found a few highly damped modes with a very small real part that
Kokkotas and Schutz were unable to find. On the other hand,
Leins {\em et al.} did not confirm the ''kink'' that occurred
for relatively high frequencies in the WKB spectra.
They suggested that this indicates that the WKB approach becomes less
reliable as the oscillation frequency increases.
This seems unreasonable: If the WKB argument is at all valid, the
approximation should become more accurate as the oscillation
frequency increases.

It is clear that the exact phase-amplitude approach outlined in the
previous section provides us with the means to test the WKB assumption
quantitatively.
By comparing the numerically determined function $q$ to the
approximation $U^{1/2}$ we can assess the reliability of
the method used by Kokkotas and Schutz. (It should be noted  that
this approximation is not exactly that used by
Kokkotas and Schutz \shortcite{kokkotas92}
where the Zerilli equation was approached directly. Nevertheless, a
study such as the present one will provide results relevant also for
that approximation.)

In approaching the problem for the stellar exterior numerically we must
first devise a scheme that avoids a problem that arises for rapidly
damped modes. The desired outgoing-wave solution to (\ref{ode}) will
grow exponentially as $r \to +\infty$. This means that it is difficult
to resolve the ingoing solution (that is exponentially small) given a
finite numerical precision. This problem is well-known from studies of
black-hole normal modes, and can be avoided by letting $r$ assume
complex values [see Andersson \shortcite{andersson92}]. Since  $\psi^-
\sim \exp(-i\omega r)$ as $r \to +\infty$ it is clear that the
exponential divergence can be suppressed along a path in the complex
$r$-plane. For large $\| r \|$ the preferred path is a straight line
with slope given by $-{\rm Im}\ \omega / {\rm Re}\ \omega$. Such
paths are  parallell to the so-called anti-Stokes lines that play a
crucial role in a complex WKB analysis \cite{andersson93}. The
anti-Stokes lines are curves along which $\int U^{1/2} dr$ is a real
quantity.  In order to deal with the present problem we will integrate
(\ref{qequation}) along a straight line (with the prescribed slope)
from a point in the asymptotic regime towards the stellar surface.
One can show, either using an analytic continuation argument or a
WKB analysis such as that of Ara\'ujo {\em et al.} \cite{nicholson},
that asymptotic conditions introduced for complex values of $r$ in the way
described above are, in fact, identical to the desired outgoing-wave boundary
condition on the real $r$-axis.

In a complex-coordinate WKB approach to (\ref{ode}) the so-called
Stokes lines also play an important role [see the discussion by
Andersson {\em et al.} \shortcite{andersson93}]. These correspond to
$\int U^{1/2} dr$ being purely imaginary. From each of the, possibly
complex, zeros of $U^{1/2}$ -- the transition points of the problem --
emanate three such contours. When an approximate solution to
(\ref{ode}) is continued across a Stokes line the character of the
solution changes. This is called the Stokes  phenomenon. Specifically
it means that, if a certain linear combination of $\psi^+$ and $\psi^-$
(with $q$ replaced by $U^{1/2}$ ) represents the desired solution to
(\ref{ode}) on one side of the Stokes line, another linear combination
should probably be used on the opposite side. Moreover, it is
straightforward to show [see the Appendix in Andersson
\shortcite{andersson93b}] that this effect is important also for
numerical integration. If the integration of (\ref{qequation}) is
continued across  a Stokes line the function $q$ may become
oscillatory.

We thus have a simple diagnostic of the ``no-reflection'' assumption
used by Kokkotas and Schutz \shortcite{kokkotas92}: If a Stokes line
(emanating from one of the complex zeros of $U^{1/2}$) crosses the real
$r$-axis between $r=R$ and $r=+\infty$ we should have
reflection.  For the stellar models considered here we find that this
does indeed happen for the new modes discovered by
Leins {\em et al.} \shortcite{leins93}. For
small oscillation frequencies a Stokes line crosses the real $r$-axis
in the relevant interval and the Stokes  phenomenon should be accounted for
in an approximate analysis. It is, of course, possible that this would
give rise to a very small correction. To test whether this is the case,
 we have compared $q$ as numerically determined from (\ref{qequation})
 to $U^{1/2}$ for several frequencies. A typical example -- for one of
 the new modes that were identified by
Leins {\em et al.} \shortcite{leins93} -- is
 displayed in Figure 1. From this Figure it follows that the WKB method
 used by Kokkotas and Schutz \shortcite{kokkotas92} cannot be trusted
 for modes with a very small real part and a relatively large imaginary
 part.

In order to illustrate the difficulties involved in numerical
integration along the real $r$-axis -- and thus the advantage of using
complex coordinates -- we also show results of a real-$r$ calculation
in Figure 1.  This specific example corresponds to the worst possible
situation. The numerically computed function $q$ has a singularity
close to the real $r$-axis. That this can be expected if one fails to
account for the Stokes phenomenon is clear from the equations given in
the Appendix of Andersson \shortcite{andersson93b}.  Note also that the
magnitude of $q$ drops dramatically immediately after passing close to
this singularity. When the surface of the star is reached $\vert q
\vert$ has decreased more than ten orders of magnitude.

\begin{figure}
\vspace{3cm}
\caption{Top: Comparing the numerically determined function $q$ (solid
line) to the WKB approximation $U^{1/2}$ (dashed line). Only the real
parts of the manifestly complex functions are shown, but the result is
similar for the imaginary parts. This example is for model 4 (see the
beginning of section 6.1 for particulars) and $\omega M = 0.35+0.84i$.
The WKB approach of Kokkotas and Schutz cannot be trusted in this case.
The numerical calculation towards the surface of the star is performed
along a straight line at an angle of roughly -67 degrees from the real
coordinate axis. This corresponds to the asymptotic direction of the
so-called anti-Stokes lines. The Stokes phenomenon gives rise to
oscillations in $q$ as one gets close to the surface. Bottom: For
comparison we show results of a similar study on the real $r$-axis. It
should be emphasized that the case shown here is one of the most
difficult, and that the agreement between the exact and the approximate
function is generally  good when ${\rm Re}\ \omega > {\rm
Im}\ \omega$. }
\end{figure}

 Our study shows that the WKB technique should be reliable for
 modes with large real parts. This conclusion would mean that the ``kink''
 in the spectrum of Kokkotas and Schutz cannot yet be dismissed. However, we
 can proceed one step further and use the numerically determined function
 $q$ in the necessary matching at the
 stellar surface. This should provide an accurate way of dealing with the
 exterior problem, and it will hopefully enable us to conclude whether
 the ``kink'' is real or not.

\section{An accurate condition for normal modes}

In principle, it is a simple task to derive a normal-mode condition
based on the phase-amplitude approach. We want to match the numerical
solution to the inside problem to the solution for the exterior at the
surface of the star. That is, we require that the Zerilli function and
its derivative be continuous at the surface $R$. Then
we can use (\ref{zpsi}) and
\begin{equation}
{dZ \over dr_\ast} = \left( 1 - {2 M \over r} \right)^{1/2}
 {d\Psi \over dr} - {M \over r^2}\left( 1 - {2M \over r}
\right)^{-1/2} \Psi \ ,
\end{equation}
which follows immediately from (\ref{tortoise}).

Let us first assume that the physically acceptable solution for
the interior of the star corresponds to a mixture of out- and
ingoing waves at spatial infinity. Then we have
\begin{equation}
\Psi = A_{\rm in} \psi^+ + A_{\rm out} \psi^- \ ,
\label{combi}\end{equation}
outside the star, and we get from (\ref{pamsols})
\begin{equation}
{d\Psi \over dr} = A_{\rm in} \psi^+ \left[ iq - {1\over 2q} {dq \over dr}
\right] -A_{\rm out} \psi^-  \left[ iq + {1\over 2q} {dq \over dr} \right]\ .
\end{equation}
The next step is to solve (\ref{qequation}) for a particular
value of $\omega$ from a point far away from the star
in the way described in the previous section. This means that the
above expressions for $\Psi$ and its derivative in terms of
$\psi^\pm$ will remain valid. Moreover, if we take
the lower limit of integration in (\ref{pamsols}) to be the surface
$R$ of the star (where we do the matching) we get
\begin{equation}
\psi^\pm (R) = q^{-1/2}(R) \ ,
\label{ampli}\end{equation}
which means that we need only determine $q$ and its derivative.

It should be pointed out that the amplitudes
$A_{\rm in}$ and $A_{\rm out}$ as defined by (\ref{combi})
differ slightly from those that follow from the asymptotic behaviour
of the Zerilli function;
\begin{equation}
Z \sim  B_{\rm in} e^{i\omega r_\ast} +B_{\rm out} e^{-i\omega r_\ast} \ .
\end{equation}
The difference is essentially a phase-factor and is of no importance for
a search for mode-frequencies. With our definition the amplitudes of
the functions that asymptotically represent out- and ingoing waves are
equal at the surface of the star [see (\ref{ampli})].

Straightforward algebra leads to
\begin{eqnarray}
&& A_{\rm in} = -{i \over 2 \sqrt{q(R)}} \left( 1 - {2 M \over  R }
\right)^{-1/2} \times \nonumber \\ && \left\{
Z_S \left[ \left( 1 - {2 M \over R} \right)
\left(iq + {1\over 2q} {dq \over dr}\right)_{r=R} + { M \over
 R^2} \right] + Z_S^\prime \right\} \ , \nonumber \\
\end{eqnarray}
and
\begin{eqnarray}
&& A_{\rm out} = {i \over 2 \sqrt{q(R)}}\left(  1 - {2 M \over R}
\right)^{-1/2} \times \nonumber \\
&& \left\{
Z_S \left[ \left( 1 - {2 M \over R} \right)
\left(-iq + {1\over 2q} {dq \over dr} \right)_{r=R} + { M \over
  R^2} \right] +Z_S^\prime \right\} \ , \nonumber \\
\end{eqnarray}
where $Z_S$ represents the interior solution at the surface of the star,
and a prime denotes a derivative with respect to $r_\ast$.
However, since the solution is only determined up to a constant
 factor by the physical conditions it makes more sense to study
\begin{equation}
{A_{\rm out} \over A_{\rm in} } = {Z_S \left[ \left( 1 - {2 M \over
R} \right) \left(iq - {1\over 2q} {dq \over dr} \right) -
 { M \over  R^2} \right] - Z_S^\prime \over  Z_S
\left[ \left( 1 - {2 M \over R} \right)
\left(iq + {1\over 2q} {dq \over dr} \right) + { M \over
R^2} \right] + Z_S^\prime } \ .
\label{ratio}\end{equation}
A quasinormal mode of the stellar system corresponds to a singularity of
this ratio as a function of $\omega$.

It may be worthwhile to clarify the meaning of $dq/dr$ in the above
equations, since we are assuming that $r$ is complex in the exterior.
In principle,  one could simply interpret (letting $z$ represent the
complex $r$) $dq/dr$ as $dq/dz$ but this may not
be very practical. As mentioned previously, we integrate for the exterior
solution along a straight line in the complex $z$-plane.
This line is parametrized by the real distance $\rho$ (from the surface
of the star) with constant phase angle $\theta$.
Then $dz/d\rho = e^{i\theta}$ and $dq/d\rho$ is the derivative
of $q$ with respect to the real parameter. Thus we get
\begin{equation}
{dq \over dr} = {dq \over dz} =  e^{-i\theta}{dq \over d\rho}
\end{equation}
which is easily implemented numerically.

\section{Discussion of numerical results}

The main purpose of the present study is to understand the
discrepancies between the results of Kokkotas and Schutz
\shortcite{kokkotas92} and those of Leins {\em et al.}
\shortcite{leins93}. Specifically, we want to find the ``new'' modes
that seem to exist for very small oscillation frequencies and large
damping \cite{leins93}. We want to see if our ad hoc explanation for
why those modes could not be found when a WKB approach to the exterior
problem was used (see section 4) is correct, {\em i.e.,} that one must
account for Stokes phenomenon for frequencies close to the imaginary
$\omega$-axis. We also feel that it is imperative that we check the
possible existence of the ``kink'' that
Kokkotas and Schutz \shortcite{kokkotas92} found in
their spectra,  especially since this feature was not
found by Leins {\em et al.} \shortcite{leins93}.

\subsection{Highly damped (w) modes}

We have performed detailed calculations for the four models studied by
Kokkotas and Schutz \shortcite{kokkotas92}. A sample of the numerical
results for model 4 (that has  characteristics: $\rho_c = 10^{16}$
g/cm$^3$, $R = 6.465$ km and $M = 1.3 M_\odot$, {\em
i.e.,} $ 2 M/ R = 0.594$) can be found in Table 1.

\begin{table}
\caption{A sample of characteristic frequencies for stellar modes
associated with  model 4 of Kokkotas and Schutz (1992). The highly
damped (w) modes are compared to the slowly damped (p) modes. The
results for w-modes are in good agreement with those of Leins {\em et
al.} (1993) The frequencies
found for the f-mode and the first of the p-modes are in perfect
agreement with those obtained by Kojima (1988).
The ``new'' w-modes found by Leins {\em et al.} are indicated by $\dag$
(the imaginary part of the one in
 parenthesis is too large for our program to provide truly reliable
 results, but there does seem to be a mode there).
The same data is shown in Figure 2. All entries are given in
 units of $M^{-1}$. The given
 damping rates for the slowly damped modes are accurate to a few parts
 in $10^7$.}
\begin{tabular}[t]{cclcll}
\multicolumn{3}{c}{Highly damped modes} & \multicolumn{3}{c}
{Slowly damped modes} \\
${\rm Re}\ \omega\ $ & ${\rm Im}\ \omega\ $ & & ${\rm Re}\ \omega\ $
& ${\rm Im}\ \omega\ $ & \\
\hline
(0.142 & 1.286) & $\dag$ & & & \\
      &       &        & 0.171 & $6.21 \times 10^{-5}$ & ($f$) \\
0.353 & 0.838 & $\dag$ & 0.344 & $2.2 \times 10^{-6}$ & ($p_1$)\\
0.471 & 0.056 &        & 0.502 & $4.05 \times 10^{-5}$ &\\
0.559 & 0.384 & $\dag$ & && \\
0.654 & 0.164 &        & 0.658 & $3.3 \times 10^{-6}$ &\\
0.892 & 0.227 &        & 0.810 & $ 5 \times 10^{-7}$ &\\
      &       &        & 0.960 & $ 4\times 10^{-7}$ & \\
1.128 & 0.262 &        & 1.100 & $ 5\times 10^{-7}$ & \\
1.363 & 0.287 &       &&&\\
1.599 & 0.307 &       &&&\\
1.836 & 0.324 &       &&&\\
2.073 & 0.339 &       & &&\\
2.310 & 0.353 &        & &&\\
2.549 & 0.365 &        & &&\\
2.788 & 0.375 &        & &&\\
\end{tabular}
\end{table}

We set out on this investigation to verify the existence of
modes with large damping and small oscillation frequencies that had been
found by Leins {\em et al.} \shortcite{leins93}, and also to find whether
there is a ``kink'' in the w-mode spectrum or not. The first of these
questions was readily answered as soon as we had combined the numerical
integration approach for the exterior vacuum (see sections 3--5) with
the numerical code that Kokkotas and Schutz had used
for the interior problem. The modes found by Leins {\em et al.}
\shortcite{leins93} do, indeed, exist. In view of our understanding of
the failure of the WKB approach for these frequencies this makes sense.
As is clear from Figure 1, the WKB approach does break down when the real
part of the frequency is small.

Leins {\em et al.} \shortcite{leins93} assumed that the ``kink'' in the
spectra of Kokkotas and Schutz \shortcite{kokkotas92} was evidence that
the WKB method fails also as  ${\rm Re}\ \omega$ increases. We have found
that this is not the case. Our first
calculations gave results in good agreement with those of
Kokkotas and Schutz for high frequencies, with a pronounced ``kink' in all
spectra, and it was clear that the WKB method is reliable for high
oscillation frequencies. Leins {\em et al.} \shortcite{leins93} argue that
their approach to the exterior problem did not involve approximations and
should therefore be considered as more reliable than the WKB
approximation. Still, we found that our new numerical scheme gave results
that did not agree well with those of Leins {\em et al.}.
At the same time, both methods have proved to be reliable and providing
results of high accuracy in the case of black holes. Hence, it seemed
likely that they should both be trusted, and  that
any discrepancies, {\em eg.}, the ``kink'', were due to differences in the
approach to the
interior problem. Unfortunately, this meant that we faced a rather
difficult situation. The approach of Detweiler and Lindblom
\shortcite{detweiler85} has been used in all  studies of highly damped
oscillations of relativistic stars. Hence, any differences
would be in the numerical codes and  rather difficult to find.

With this in mind we scrutinized our code for the interior problem, and
found points where it could be improved. The amended code is
much more reliable, and numerically accurate, than the original one used by
Kokkotas and Schutz. It also
runs considerably faster. It turns out that most of the improvements have a
minor effect on the numerical results, however. (The results quoted in
Table 1 were, obviously, computed with the final version.) After testing the
code we have a much clearer idea of its restrictions. Most importantly,
we have found that the ``kink'' is intimately related to the choice of
point close to the centre of the star where power series expansions are
used to initiate the numerical integration.
When choosing a starting point one introduces an artificial length-scale
in the problem, and this scale is reflected by the oscillation frequency at
which the ``kink'' occurs. This can easily be illustrated if one considers
the eigenfunctions: As the oscillation frequency increases the number of
nodes in each eigenfunction increases. One will eventually reach a
frequency where nodes should occur for smaller radii than the one where
the numerical integration is initiated. The failure of the power-series
solutions to account for this feature of the physical solutions leads
to the ``kink'' in the spectrum.   By initiating the integration
sufficiently close to the centre of the star one can
ensure that this ``kink'' does not occur in the frequency regime of interest.
In the calculations discussed here we initiated the integration at a point
roughly corresponding to $2.5 \%$ of $R$. It seems likely that
Leins {\em et al.} \shortcite{leins93}
initiated their integration at a point closer to the centre than we did
originally, and had they continued their calculation to higher oscillation
frequencies they would have found a ``kink'' similar to that of
Kokkotas and Schutz \shortcite{kokkotas92}.
The ``kinks'' are numerical artefacts.

We also find that it
is extremely difficult to avoid numerical noise in the eigenfunctions.
This is mainly due to the fact that the eigenfunction $X$ decreases
rapidly towards the surface of the star (where it should vanish).
When solving for the interior of the star one matches a solution from
the centre (that satisfies the physical condition of regularity) to one
from the surface of the star at some intermediate point. In
order to get a reliable result it is imperative that this matching is not
done at too large a radius. We find that the interior matching should be
performed for radii smaller than something like $0.25 R$.
Furthermore, the calculation gets more
difficult as ${\rm Im}\ \omega$ increases. This is due to the fact that
$\| X \|$ is then many magnitudes smaller than any of the other three
eigenfunctions $K, H_1$ and $W$. Thus it is difficult to compute $X$
with acceptable accuracy. For this reason the calculation is more
difficult for  less relativistic stellar models: The damping of the
w-modes generally increases as the model gets less relativistic.
Interestingly, this explains why the results of
Kokkotas and Schutz \shortcite{kokkotas92} for model 1
were not as good as for the
other three models. In view of this fact we decided to restrict our
detailed study to  models 2--4 of Kokkotas and Schutz
\shortcite{kokkotas92}, and also not
 to attempt a search for modes with ${\rm Im}\ \omega M > 1$. We
feel that the interior problem must be reformulated if a reliable
search for modes with very large damping is to be made. In this
context a study based on the alternative formulations of Chandrasekhar
and Ferrari \shortcite{chandra91a}, or Ipser and Price
\shortcite{ipser91}, may provide interesting results. However, this
does not mean that our results should not be taken seriously.
We believe that the present study is the most accurate one to
date within the range of validity that we have indicated.

\subsection{Slowly damped (p) modes}

An interesting effect of the improvements that we made to our code for
the interior problem is that it is now considerably more accurate for modes
with very slow damping. Hence, it is not at all difficult to compute
the f-mode for the various models. Furthermore, we are able to iterate
 for several of the first p-modes. As far as we know noone has
attempted to compute these modes before, and the results that we obtain
are somewhat surprising. The damping of the p-modes does not necessarily, as
has been generally presumed, decrease rapidly as one proceeds to higher
oscillation frequencies. In fact, for the most compact of our models
(model 4) we find an intriguing behaviour. There is a low-order mode that
damps away much faster than expected (see Table 1).
This more rapidly damped mode occurs only for the most compact of our
stellar models.

\begin{figure}
\vspace{3cm}
\caption{The p- and w-mode spectra for model 4, cf. Table 1.
Note the absence of a ``kink'' for high oscillation frequencies in the
w-mode spectrum, and that the damping does not decrease monotonically as
one proceeds up the p-mode spectrum.  }

\end{figure}

This is an interesting result that could be of potential importance for
many of our astrophysical expectations, such as the life-time of
gravitational-wave sources. Furthermore,  methods used by other
authors to study f-modes could be used to test its
correctness. For model 4 the ratio of real to imaginary part for the
relevant mode is not considerably different from that for the
f-mode. Therefore, it seems likely that the program used by Detweiler
and Lindblom \shortcite{detweiler85}, or that of Kojima
\shortcite{kojima88}, could be used to test our results.
In fact, Dr Kojima and Dr Lindblom have both very kindly performed
this calculation. The results they obtain for the p-modes of model 4 are
in good agreement with those listed in our Table 1.

It may be worthwhile to comment on the difficulty of finding p-modes
here. It is clear that a numerical code must be very robust and
accurate if iteration for these slowly damped modes is to be at all
possible. The imaginary part is typically at least six orders of
magnitude smaller than the real part, and if one requires single-digit
precision in the imaginary part one must therefore achieve at least six
digits in the real part. Our amended code should be reliable for
imaginary parts larger than $10^{-7}$.

In cases when the imaginary part of the mode-frequency is so small that
iteration is not possible, one can always attempt to infer the real
part from a graphical approach. We know that the modes correspond to
the zeros of the asymptotic amplitude $A_{\rm in}$ (or the
singularities of the ratio $A_{\rm out} /A_{\rm in}$). If such a zero
is situated close to the real $\omega$-axis it should, in principle,
be easy to distinguish in a plot of $\log \| A_{\rm in} \|$. This idea
was recently exploited by, for example,
Ferrari and Germano \shortcite{ferrari94}.
That it works nicely is illustrated in Figure 3.
When generating this figure we  normalised the perturbed quantities
in such a way that the radial displacement $W$ is equal to 1 at the surface
of the star for all frequencies. The slowly damped
p-modes give rise to very narrow singularities that are easy to
distinguish although the data in the figure correspond to  $\Delta
\omega M = 10^{-3}$. It is not very time consuming to create a figure
with this kind of resolution,  and it is certainly worthwhile. In fact,
we can not only pin down all the p-modes given in Table 1, but also
guess at which oscillation frequencies there will be a w-mode. It seems
as if a peak in $\log \| A_{\rm in} \|$ is an indication of a w-mode
higher up in the complex $\omega$-plane. We did not expect to find this
feature in this kind of graph and the exact origin of these peaks is
something of a mystery. Interestingly the peaks do not show up in similar
graphs generated by the numerical code of Lindblom (private communication).
It thus seems that the difference between the two approaches, {\em i.e.}
between (i) matching the interior solution to the exterior one at the surface
of the star and (ii) using the value of the Zerilli function at the
surface as initial data for integration towards infinity where the
solution is matched to an asymptotic form, is crucial.
We believe that the information contained in this sort of graph is important
to our understanding of the stellar pulsations and are presently investigating
this issue further. Although we do not yet have the a clearcut explanation of
the phenomenon it is worthwhile to make one final point.
It is easy to show that the peaks in Figure 3 originate from the interior
solution and not the exterior one. Remembering that the Zerilli function is
generated from the two spacetime perturbations $K$ and $H_1$, this agrees
well with the notion that the spacetime perturbations dominate the fluid
ones (which set the normalisation for the figure) at the w-mode frequencies.

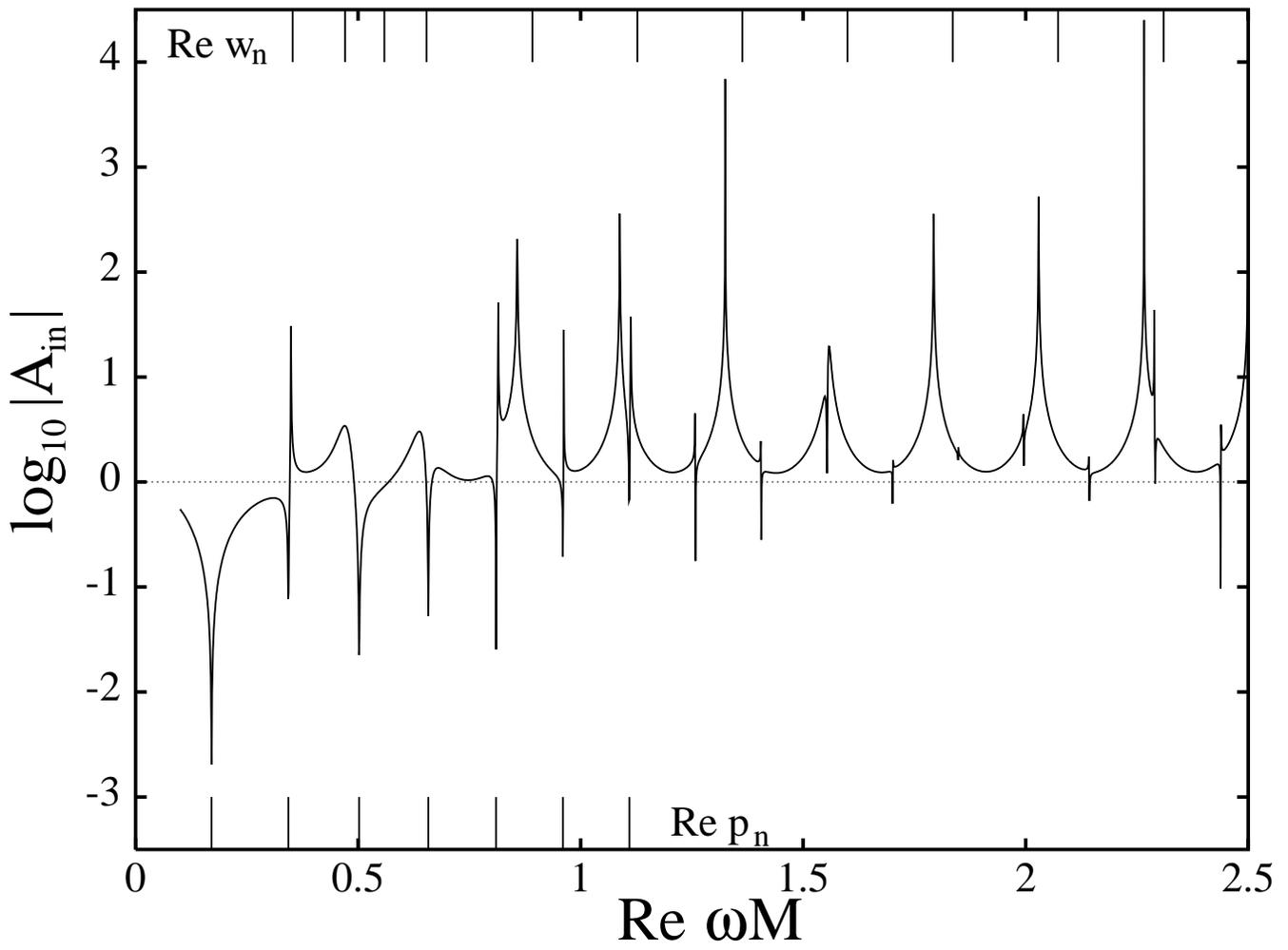
\begin{figure}
\vspace{3cm}
\caption{Illustration of the graphic approach for finding p-modes. We
show $\log_{10} \| A_{\rm in} \|$ as a function of real $\omega M$ for
model 4.  A slowly damped mode corresponds to a narrow singularity, and
all the p-modes listed in Table 1  can easily be distinguished. Peaks
in this figure  indicate the existence of the w-modes.
The real parts of the w and p-modes are shown as vertical lines at the
top and bottom of the figure, respectively.}
\end{figure}

\subsection{Are there different families of highly damped modes?}

Leins {\em et al.} \shortcite{leins93} argued that they had found a new
family of highly damped modes. Given the present evidence we have to
assess whether that is the case or not. We have shown why Kokkotas and
Schutz \shortcite{kokkotas92} could not find modes with high damping
and small oscillation frequencies with the WKB scheme
they used. But surely, that in itself does not imply that any modes
they could not find belongs to a new family? Different families of
modes must be distinguished for clear physical reasons. The difference
between the two families (p- and g-modes) of slowly damped modes is
well established, and the w-modes are certainly distinct. But given the
knowledge available at the moment it is not at all clear that a similar
split of the highly damped modes into different classes makes sense.

Leins {\em et al.} \shortcite{leins93} put forward two arguments for
why their new modes belong to a different family. The first  relates to
the number of
 nodes of the eigenfunctions corresponding to the various modes. They
find that the number of nodes increase systematically as one progresses
up the w-mode sequence, but that the situation is not that clear for
the new modes. In doing this they claim to be studying the
``amplitude'' of the eigenfunctions. What they actually study is the
number of nodes in the real part of each eigenfunction.
An argument solely based on the real part of a manifestly complex
eigenfunction clearly does not make much sense. From, for example,
(\ref{metric})
it is evident that both the real and the imaginary part will be
relevant. Hence, the nodes in the imaginary part of the eigenfunction
must also be studied. When that is done the conclusion regarding the
higher oscillation frequencies remains the same, but for the new modes
the situation is somewhat
clearer. The imaginary part of each eigenfunction may have a node even
if the real part does not, and vice versa.
However, although it may provide a useful diagnostic, the number of nodes
in the eigenfunctions is not a very good measure for distinguishing different
families of modes. It is well known [see for example chapter 17 in Cox
\shortcite{cox}] that different p-modes may have an identical number of
nodes in the eigenfunctions, and there is no reason why that could not
be the case also for highly damped modes.

However, Leins {\em et al.} \shortcite{leins93}  argue that their new modes
are different from the ones of Kokkotas and Schutz
\shortcite{kokkotas92} because of other features of the eigenfunctions.
For the new modes the eigenfunctions die out towards the centre of the
star, whereas the amplitude stay roughly constant  for all values of
$r$ for the old modes.
This difference is similar to that between g- and p-modes \cite{cox}.
For the g-modes -- which correspond to oscillation frequencies smaller
than that of the f-mode -- the eigenfunctions are known to die out
towards the centre of the star, whereas this is not the case for the
p-mode eigenfunctions. Could it be that the perturbations generally die
out towards the centre of the star for very small oscillation
frequencies? If so, a distinction of different w-mode families based
on this feature would not be satisfactory.

At the present time, it seems premature to divide the w-modes of
relativistic stars into different ``families''. If such a classification is
to make sense it must be based on clear physical principles, and our
understanding of the highly damped stellar oscillations is still far
from satisfactory.
The issue could probably be resolved by further detailed studies of
different families of stellar models \cite{leins93}.
A vital piece of information that is still missing
from the puzzle regards the possible existence of modes with very large
imaginary parts. As mentioned previously we have found that the present
approach to the interior problem suffers from numerical difficulties
when ${\rm Im}\ \omega M > 1$ or so. This is not to be taken as an
indication that
modes with larger imaginary parts do not exist. On the contrary, such
modes may well exist and a mode survey that covers the entire
complex frequency plane could produce interesting
results.

\section{Concluding remarks}

In order to understand the discrepancies between results obtained by
Kokkotas and Schutz \shortcite{kokkotas92} and Leins, Nollert and
Soffel \shortcite{leins93}, we have computed highly damped modes for
relativistic stars. The stellar models considered are simple polytropes
used in several previous investigations. We have outlined a new
numerical approach to the problem for the exterior of the star. This
new scheme is based on numerical integration for complex coordinates,
and is similar to one that has proved extremely reliable for black-hole
problems \cite{andersson92}.

With this new approach to the stellar exterior we are able to find
modes that were not identified by Kokkotas and Schutz
\shortcite{kokkotas92}. These modes, that are highly damped and
situated close to the imaginary frequency-axis, agree perfectly with
modes found by Leins {\em et al.} \shortcite{leins93}. That the WKB
method employed by Kokkotas and Schutz failed to distinguish these
modes can be understood if the so-called Stokes phenomenon (familiar
from WKB theory) is accounted for.

We have managed to explain the occurence of a ``kink'' in the spectra of
Kokkotas and Schutz: It is a numerical artefact due to the choice of point
close to the centre of the star where power-series expansions are used to
initiate the numerical integration. This conclusion was drawn after we
had scrutinized the way that we dealt with the
stellar interior. We also found  other points where our original code could
be improved. Although none of those changes affected the numerical results
significantly, the program is now more robust
and runs considerably faster. That it allows us to do  accurate
calculations is illustrated by the fact that we could iterate for several
of the very slowly damped p-modes. This led to a rather surprising result:
For the most compact of our stellar models (corresponding to
  $ 2 M/ R = 0.594$)  the damping of the p-modes does not decrease
monotonically as one proceeds to higher oscillation frequencies.
There is a low-order mode that  damps away at least ten times  faster than
anticipated. This result may significantly affect our expectations regarding,
for example, stellar
stability and the lifetime of gravitational-wave sources. Hence,
it is of some importance that it be studied in more detail.

The new complex-coordinate method that we have employed in the
present study seems promising also for rotating stars, {\em i.e.}, when
the exterior spacetime is only known approximately. Since all previously
suggested methods will surely fail to handle that difficulty this new
approach could turn out to be of considerable importance. The anomalous
p-mode damping that we have discovered here would be especially interesting
if it exists for rotating star f-modes, because these limit stability.
Such issues are, however,  beyond the scope of the present investigation
and we will return to them in the future.

\section*{Acknowledgements}

We would like to thank Hans-Peter Nollert for providing us with  numerical
results that were not published in their paper. Discussions with him provided
valuable information that helped us find the reason for the discrepancies
between the previous studies of this problem. We are also grateful to
Yasufumi Kojima and Lee Lindblom who most kindly provided us with
independent checks of  our results for slowly damped modes.

\end{document}